\documentclass[10pt,journal,compsoc]{IEEEtran}
\ifCLASSOPTIONcompsoc
  \usepackage[nocompress]{cite}
\else
  \usepackage{cite}
\fi
\ifCLASSINFOpdf
\else
\fi
\usepackage{amsmath}
\usepackage{amsthm}
\usepackage{amsfonts}
\usepackage{todonotes}

\newtheorem{deft}{Definition}

\usepackage{url}

\begin{document}
%
\title{A Neural Network Based Method with Transfer Learning for Genetic Data Analysis}
%
%
%
%

\author{Jinghang Lin,
        Shan Zhang, Qing Lu*
\IEEEcompsocitemizethanks{
\IEEEcompsocthanksitem Jinghang Lin is with Department of Biostatistics, Yale University, Connecticut, U.S.A.
Email: jinghang.lin@yale.edu.
\IEEEcompsocthanksitem Shan Zhang is with Department of Statistics and Probability, Michigan State University, Michigan
U.S.A.
\IEEEcompsocthanksitem Qing Lu is with Department of Biostatistics , University of Florida, Florida
U.S.A.
\protect\\

}
\thanks{Manuscript received April 19, 2005; revised August 26, 2015.}}

%
%

\markboth{Journal of \LaTeX\ Class Files,~Vol.~14, No.~8, August~2015}%
{Shell \MakeLowercase{\textit{et al.}}: Bare Demo of IEEEtran.cls for Computer Society Journals}
%



\IEEEtitleabstractindextext{%
\begin{abstract}
Transfer learning has emerged as a powerful technique in many application problems, such as computer vision and natural language processing. However, this technique is largely ignored in application to genetic data analysis. In this paper, we combine transfer learning technique with a neural network based method(expectile neural networks). With transfer learning, instead of starting the learning process from scratch, we start from one task that have been learned when solving a different task. We leverage previous learnings and avoid starting from scratch to improve the model performance by passing information gained in different but related task. To demonstrate the performance, we run two real data sets. By using transfer learning algorithm, the performance of expectile neural networks is improved compared to expectile neural network without using transfer learning technique.
\end{abstract}

}

\maketitle

\IEEEdisplaynontitleabstractindextext

%
\IEEEpeerreviewmaketitle

\IEEEraisesectionheading{\section{Introduction}\label{sec:introduction}}

%
%
%
%
\IEEEPARstart{T}{raditional}
machine learning models focus on one single and specific task. If we have two related tasks, one task could inherit some information from the other task. It is natural to store knowledge gained while solving one problem and applying it to a different but related problem. For example, knowledge gained while learning to recognize cats could apply when trying to recognize dogs for image classification problem. We call this technique as transfer learning. The insight of transfer learning is motivated by the fact that human can intelligently apply knowledge learned previously to solve new problems
efficiently. It is easier to transfer knowledge if two tasks are more related.

With the wide application, transfer learning has become a popular and promising area in machine learning. For example, transfer learning is a popular method in computer vision because it allows us to build accurate models in a more efficient way\cite{RWW,MSM,KTK}. Transfer learning has also been implemented in natural language processing (NLP)\cite{CRN}, medical image\cite{KAB}. Some works are worth to be mentioned. Syed proposes seeded transfer learning in a 
regression context to improve prediction performance in target domain\cite{SMS}. Yosinski et al. show that how lower layers in neural networks act as conventional computer-vision feature extractors, such as edge detectors, while the final layer works toward task-specific features\cite{JYJ}. Rosenstein uses naive Bayes classification algorithm to detect, perhaps implicitly, that the inductive bias learned from the auxiliary tasks will actually hurt performance on the target task \cite{MRZ}.

However, little attention of transfer learning research  has been attracted to multi-modal biomedical data, such as genetic data. 
In this paper, we focused on applying transfer learning technique into a neural network based method(expectile neural networks) to give a prescriptive and predictive analytics based on genetic sequencing data. We focus on parameter transfer or instance reweighting. This approach works on the assumption that the models for related tasks share some parameters. There are some advantages of doing these. First, if source task and target task are relevant, we could improve our result. Second, since we inherit some parameters from initial task, the number of parameters in target task are reduced which give us some computational advantages especially in large data set.

The paper is organized as follows. Section 2 introduces the expectile neural networks and transfer learning.
In section 3, expectile neural networks with transfer learning are implemented in two real data sets. Section 4 gives summary and discussion.
\section{Method}
Expectile regression is first proposed by Newey and Powell as a generalization of linear regression\cite{ALSE}. It adopts asymmetric least squares as loss function, which provides a convenient and relatively efficient approach to summarize the conditional distribution. It shows some advantages over linear regression under heteroscedastic and asymmetric scenarios.

By integrating the idea of expectile regression with neural networks,  expectile neural networks (ENN) is proposed to model the complex relationship between genotypes and phenotypes\cite{JHL}. We briefly introduce the expectile neural network in this section.  Suppose we have $n$ samples $\lbrace (\mathbf{x_i},y_i),i=1,...,n \rbrace$, where $\mathbf{x_i}=(1,x_{i,1},...,x_{i,p})^T$. $y_i$  is the phenotype for $i$th sample which has denote $p-$dimensional covarites. For example, $y_i$ could be the type of diabetes or height of one person. The covariates are mainly genetic variants, such as single nucleotide polymorphisms (SNPs), which are typically coded according to the number of minor frequent allele (e.g., AA=2, Aa=1, aa=0). To increase prediction performance, the covariates $\mathbf{x_i}$ could also include demographic characteristics (e.g., gender, age). 

\subsection{Expectile neural networks}
The major difference between expectile neural networks and classical neural network is the loss function. Classical neural network adopts $L_2$ squared loss function, while expectile neural networks(ENN) uses asymmetric $L_2$ squared loss function. An asymmetry coefficient $\tau$ is given in loss function of ENN. the $\tau-$mean quantifies different 'locations' of a distribution, and thus it can be viewed as a generalization of the mean and an alternative measure of 'location'
of a distribution\cite{GYZ}. If $\tau = 0.5$, ENN degenerates to a classical neural network. Therefore, ENN can also be viewed as a
generalization of a classical neural network. In ENN, we don't assume a particular functional form of covariates and use neural networks to approximate the underlying expectile regression function. In order to model a complex relationship between covariates and phenotypes, we integrate the idea of neural networks with expectile regression. We illustrate ENN with one hidden layer in Fig 2. By adding more hidden layers, ENN method can be easily extended to a version of deep expectile neural network. We give the definition of ENN proposed by Lin et al\cite{JHL} .

Given the $\mathbf{x}_t$, we first build the first layer of hidden nodes $h_{q,t}$,
\begin{equation}\label{hiddenlayer}
h_{q,t}= f^{(1)}(\sum_{p=1}^{P} x_{p,t}w_{pq}^{(1)}+b_{q}^{(1)}),  q= 1,...,Q, t = 1, ..., n.
\end{equation}

where $w_{pq}$ denotes weights and $b_{q}$ denotes the bias;  $f^{(1)}$ is the activation function for the hidden layer that can be a sigmoid function, a hyperbolic tangent function, or a rectified linear units(ReLU) function. Similar to hidden nodes in neural networks, the hidden nodes in ENN can learn complex features from covariates $\mathbf{x}$, which makes ENN capable of modelling non-linear and non-additive effects. Based on these hidden nodes, we can model the conditional $\tau$-expectile, $\hat{y}_{\tau}(t)$,
\begin{equation}\label{output}
\hat{y}_{\tau}(t)=f^{(2)}(\sum_{q=1}^{Q} h_{q,t}w_{q}^{(2)}+b^{(2)}),
\end{equation}
where $f^{(2)}$, $w_{q}^{(2)}$, and $b^{(2)}$ are the activation function, weights, and bias in the output layer, respectively. $f^{(2)}$ can be identity function, sigmoid function, or a rectified linear units(ReLU) function.

From equations (\ref{hiddenlayer}) and (\ref{output}), we can have the overall function $f$:
\begin{equation}
f=f^{(2)}(\sum_{q=1}^{Q} f^{(1)}(\sum_{p=1}^{P} x_{p,t}w_{pq}^{(1)}+b_{q}^{(1)})w_{q}^{(2)}+b^{(2)}). 
\end{equation}
Then $\hat{y}_{\tau}(t) = f(\mathbf{x_i}).$
To estimate $w_{pq}^{(1)}, b^{(1)}_{q}, w_{q}^{(2)}, b^{(2)}$, we minimize the empirical risk function,

\begin{equation}
\mathcal{R}(\tau)=\frac{1}{n}\sum_{i=1}^{n} L_{\tau}(y_i,f(\mathbf{x_i})),
\end{equation}
where
\begin{equation}
L_{\tau}(y_{i}, f(\mathbf{x_i}))=\left\{
\begin{aligned}
&(1- \tau) (y_{i} - f(\mathbf{x_i}))^2, & if \ y_{i} < f(\mathbf{x_i}) \\
&\tau (y_i - f(\mathbf{x_i})))^2, & if \ y_i \geq f(\mathbf{x_i}).
\end{aligned}
\right.
\end{equation}
The model tends to be overfitted with the increasing number of covariates.  To address the overfitting issue, a $L_2$ penalty is added to the risk function,
\begin{equation}\label{loss}
\mathcal{R}(\tau)=\frac{1}{n}\sum_{i=1}^{n} L_{\tau}(y_i,f(\mathbf{x_i}))+ \lambda\sum_{p=1}^{P} \sum_{q=1}^{Q}\left((w_{pq}^{(1)})^2 + (w_{q}^{(2)})^2 \right)^2.
\end{equation}

The loss function for ENN is differentiable everywhere, and therefore we can obtain the estimator of expectile neural network by using gradient-based optimization algorithms (e.g.,quasi-Newton Broyden–Fletcher–Goldfarb–Shanno (BFGS) optimization algorithm).

\subsection{Transfer learning}
For the definition of transfer learning, we follow the survey by Pan and Yang \cite{SJ}. For simplicity, we only consider the scenario where there is only one source domain $\mathcal{D}_S$ and one target domain $\mathcal{D}_{T}$ s which is the most popular of the research works in the literature. Compared to traditional machine learning techniques which normally train from scratch, transfer learning techniques aim to transfer the knowledge from source tasks to a target task. A graphical illustration is shown in Fig.1.
\begin{deft}
Given a source domain $\mathcal{D}_{S}$ and learning task $\mathcal{T}_{S}$, a target domain$\mathcal{D}_T$ and learning task $\mathcal{T}_{T}$, transfer learning aims to help improve the learning of the target predictive function $f_{T}(\cdot)$ in $\mathcal{D}_{T}$ using the knowledge in $\mathcal{D}_{S}$ and $\mathcal{T}_{S}$, where $\mathcal{D}_{S} \neq \mathcal{D}_{T}$ or $\mathcal{T}_{S} \neq \mathcal{T}_{T}$.
\end{deft}
Based on different conditions for differences between source domain and target domain and differences between source task and target task, transfer learning
scenarios can be categorized differently: homogeneous transfer learning and heterogeneous transfer learning\cite{TL2014}. In this paper, we consider heterogeneous transfer learning with parameter transfer which means $\mathcal{D}_{S} =  \mathcal{D}_{T}$, $\mathcal{T}_{S} \neq \mathcal{T}_{T}$.

To have a better understanding of transfer learning, a graphical representation of ENN with transfer learning with one hidden layer is given in Fig 2. This method can be easily extended to deep ENN with multiple layers. The same input($\mathcal{D}_{S} =  \mathcal{D}_{T}$) 
in model 1 and model 2 are SNPs with 3 types of value: 0, 1, 2. The responses in model
1 and model 2 ($\mathcal{T}_{S} \neq  \mathcal{T}_{T}$) are different but related. We
tend to transfer parameters from input layer to output layer learned in model 1 to 
model 2. To achieve the optimal performance improvement for a target domain given a 
source domain, we try different source and target domains. Then we discover 
transferable knowledge to improve transfer learning effectiveness. 
\begin{figure}
    \begin{center}
      \includegraphics[scale = 0.45]{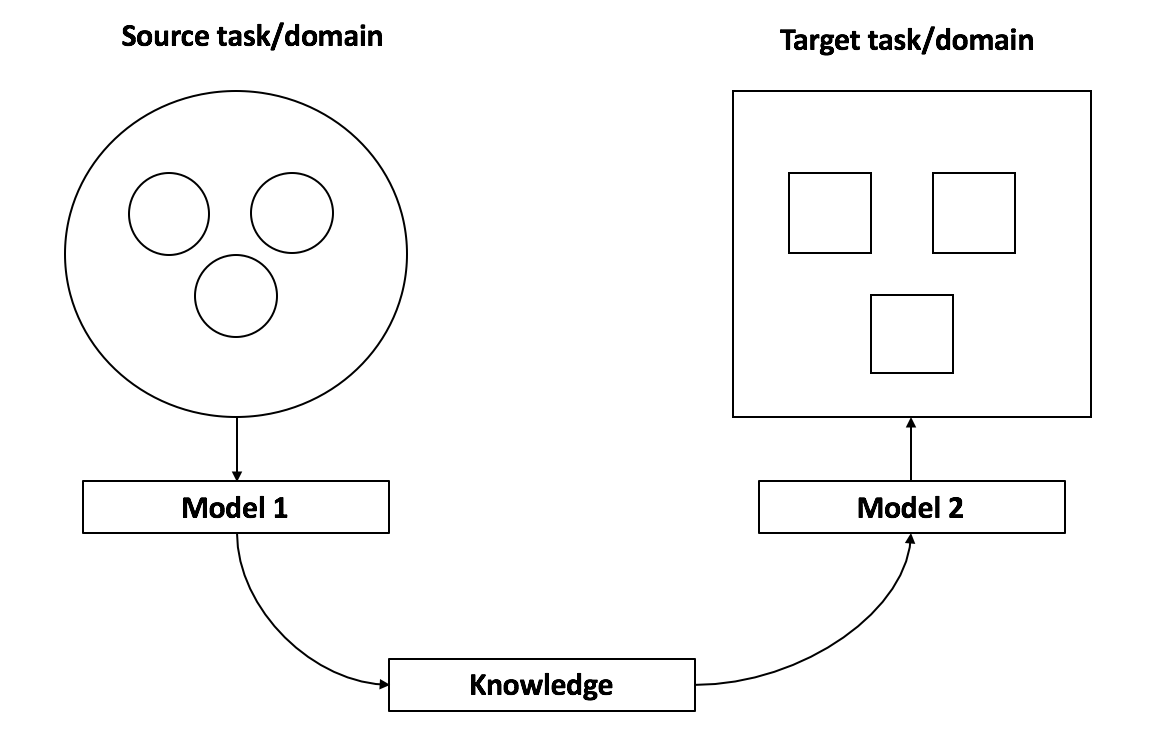}
    \caption{A illustration of transfer learning from source task to target task}
    \label{fig:my_label}  
    \end{center}
\end{figure}
\begin{figure}
    \centering
    \includegraphics[scale = 0.45]{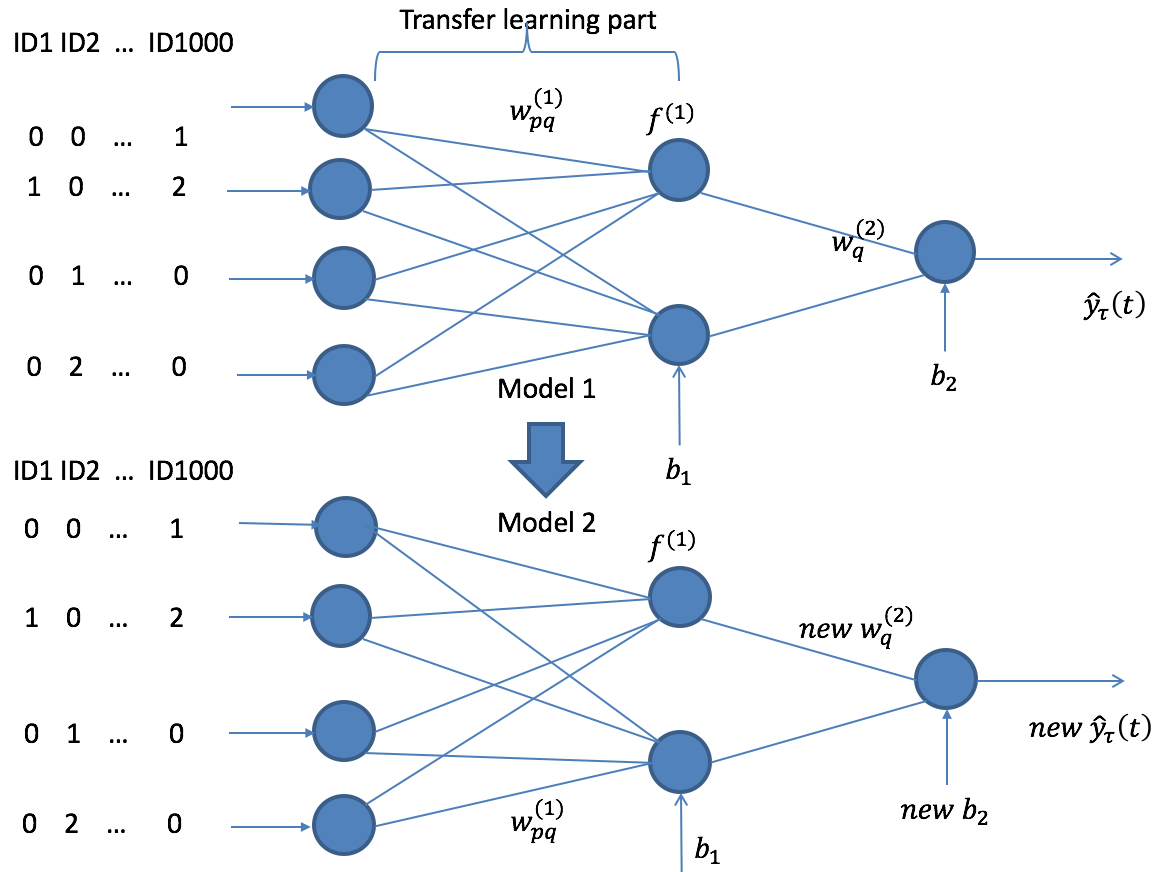}
    \caption{A illustration of transfer learning in expectile neural networks}
    \label{fig:my_label}
\end{figure}

\section{Real data application}

In this section, we integrated ENN with transfer 
learning technique to improve prediction performance. To verify if
transfer learning technique works, we ran two real data sets to 
compare the performance of ENN with transfer learning and ENN without 
transfer learning. We divided the samples into training, 
validation, and testing sets with the ratio 3: 1: 1. Even if a 
variety of activation functions can be used in ENN, such as sigmoid and tanh function, we chose 
the Rectified Linear Unit (ReLU) due to its relative performance 
and computational advantage\cite{DL}. ENN discovered different 
transferable knowledge, and thereby led to uneven transfer 
learning effectiveness which was evaluated by the performance 
improvement over non-transfer baselines in a target domain. This 
final model was then evaluated on the testing set by using the 
mean squared error ($MSE = \sum_{i=1}^{n}(y_i -\hat{y}_i)^2$)
\subsection{Real data application I}
 Alcohol consumption and tobacco use are closely linked 
 behaviors. In other words, people who drink alcohol are more 
 likely to smoke (and vice versa) and people who drink larger 
 amounts of alcohol tend to smoke more cigarettes. Data shows 
 that people who are dependent on alcohol are three times more 
 likely then those in the general population to be smokers, and 
 people who are dependent on
tobacco are four times more likely than the general population
to be dependent on alcohol\cite{EWE}.  Although tobacco and 
nicotine have very different effects and mechanisms of action, 
they might act on common mechanisms in the brain, creating 
complex interactions\cite{FDM}.
 The importance of genetic influences on both alcoholism and 
 smoking has gained widespread recognition over the past decade. 
 Several research work have indicated that a substantial shared 
 genetic risk exists between smoking and alcoholism — that is, 
 genetic factors that increase the risk for smoking also increase
 the risk for alcoholism and vice versa\cite{KVB,PKA}. Therefore,
 it is worthwhile to study two different but related tasks: 
 alcohol-related phenotype and tobacco use-related phenotype.

We applied ENN to the genetic data from the Study of Addiction: Genetics and Environment(SAGE). The participants of the SAGE are selected from three large, complementary studies: the Family Study of Cocaine Dependence(FSCD), the Collaborative study on the Genetics of Alcoholism(COGA), and the Collaborative Genetic Study of Nicotine Dependence(COGEND). We varied $\tau$ values (i.e., $0.1, 0.25, 0.5, 0.75, 0.9$) and compared ENN and ENN with transfer learning based on MSE.

The response is smoking quantity which is measured by largest number of cigarettes smoked in 24 hours, ranged from 0-240. And the drinking quantity is measured by largest number of alcoholic drinks consumed in 24 hours, range from 0-258.  We only included 3888 Caucasian and African American samples. After quality control, 149 SNPs remained for the analysis. To have better performance, we transfer smoking-related information to drinking-related information. We use the following algorithm. First, we choose smoking quantity as phenotype, and get the estimator of expectile neural network in the training data set. 
Second, we get the estimator obtained from first step as initial value(transfer learning part).
Third, we choose drinking quantity as new phenotype and keep the parameter from input layer to hidden layer and then train expectile neural network again. 
Finally, we compare two models: ENN with transfer learning technique and  ENN without transfer learning technique to demonstrate if transferable knowledge is suitable.

We divide the data into three parts randomly with 50 replicates : training, validation, testing with ratio 3:1:1. To choose the best $\lambda$, we used the grid search with different values of 0, 0.1, 1, 10 and 100. We used 1000 epochs to train the ENN model and chose 3-10 hidden units based on simulated scenarios. To reduce computation burden, we did not use a large number of hidden units. The number of hidden units is chosen to ensure that the performance of the ENN model is reasonable well. 

\begin{table}[h]
\caption{The accuracy performance of two models built by ENN with transfer learning(ENN.TF) and ENN in gene CHRNA5} \label{comparsion with two models}
\begin{center}
\begin{tabular}{lllllll}
\hline
       & \multicolumn{2}{c}{\textbf{ENN.TF}}                                       &  & \multicolumn{2}{c}{\textbf{ENN}}    &  \\ \cline{1-3} \cline{5-6}
$\tau$ & \multicolumn{1}{c}{\textbf{Train}} & \multicolumn{1}{c}{\textbf{Test}} &  & \multicolumn{1}{c}{\textbf{Train}} & \multicolumn{1}{c}{\textbf{Test}} &  \\ \hline
0.1    & 551.83      &605.79    &     &546.90           &672.44         \\ \hline
0.25   & 325.84      &439.18    &      &321.94         &473.10          \\ \hline
0.5    & 282.57       &433.06  &     &275.83           &444.16         \\ \hline
0.75   & 304.81      &484.60   &        &297.81          &487.43        \\ \hline
0.9    & 347.17       &544.24   &        &339.79          &549.08         \\ \hline
\end{tabular}
\end{center}
\end{table}

\begin{figure}
    \centering
    \includegraphics[scale = 0.5]{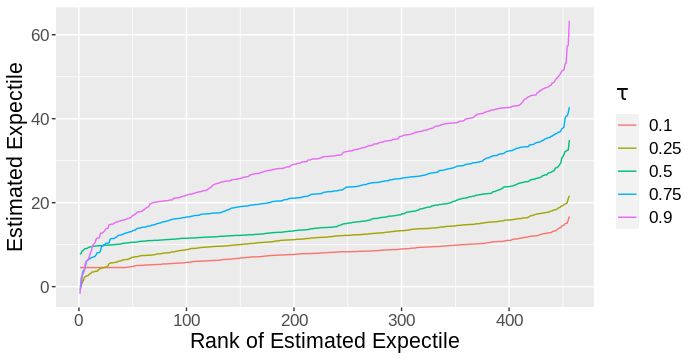}
    \caption{The conditional distribution of smoking quantity for five expectile levels (i.e., 0.1, 0.25, 0.5, 0.75, and 0.9) in gene CHRNA5}
    \label{fig:my_label}
\end{figure}

\begin{table}[h]
\caption{The accuracy performance of two models built by ENN with transfer learning(ENN.TF) and ENN in gene CHRNA3} \label{comparsion with two models}
\begin{center}
\begin{tabular}{lllllll}
\hline
       & \multicolumn{2}{c}{\textbf{ENN.TF}}                                       &  & \multicolumn{2}{c}{\textbf{ENN}}    &  \\ \cline{1-3} \cline{5-6}
$\tau$ & \multicolumn{1}{c}{\textbf{Train}} & \multicolumn{1}{c}{\textbf{Test}} &  & \multicolumn{1}{c}{\textbf{Train}} & \multicolumn{1}{c}{\textbf{Test}} &  \\ \hline
0.1    &554.11      &605.09   &      &533.04           &753.96          \\ \hline
0.25   &325.71       &441.47  &      &311.85           &517.05         \\ \hline
0.5    &281.20       &439.40   &     &260.44           &491.62        \\ \hline
0.75   &304.60       &486.80   &      &292.63          &502.86         \\ \hline
0.9    &350.01       &558.95    &     &335.89          &573.92        \\ \hline
\end{tabular}
\end{center}
\end{table}
\begin{figure}
    \centering
    \includegraphics[scale=0.5]{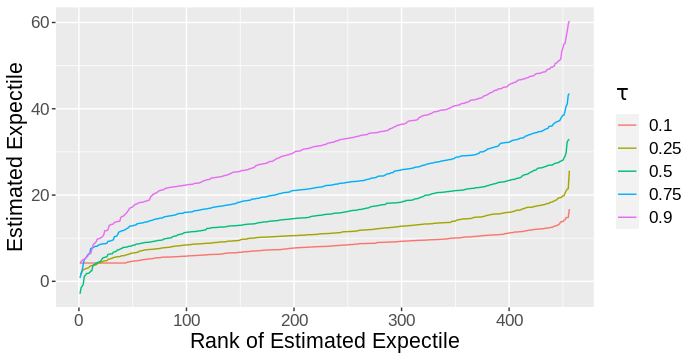}
    \caption{The conditional distribution of smoking quantity for five expectile levels (i.e., 0.1, 0.25, 0.5, 0.75, and 0.9) in gene CHRNA3}
    \label{fig:my_label}
\end{figure}

\begin{table}[h]
\caption{The accuracy performance of two models built by ENN with transfer learning(ENN.TF) and ENN in gene CHRNB4} \label{comparsion with two models}
\begin{center}
\begin{tabular}{lllllll}
\hline
       & \multicolumn{2}{c}{\textbf{ENN.TF}}                                       &  & \multicolumn{2}{c}{\textbf{ENN}}    &  \\ \cline{1-3} \cline{5-6}
$\tau$ & \multicolumn{1}{c}{\textbf{Train}} & \multicolumn{1}{c}{\textbf{Test}} &  & \multicolumn{1}{c}{\textbf{Train}} & \multicolumn{1}{c}{\textbf{Test}} &  \\ \hline
0.1    &558.39      &622.18    &     &564.57          &673.97          \\ \hline
0.25   &327.63       &448.63   &     &325.48          &473.34         \\ \hline
0.5    &283.28       &435.11   &     &270.50           &453.76        \\ \hline
0.75   &306.05      &488.15   &      &303.02          &489.71         \\ \hline
0.9    &349.24       &544.85   &      &343.52          &553.41       \\ \hline
\end{tabular}
\end{center}
\end{table}
\begin{figure}
    \centering
    \includegraphics[scale=0.5]{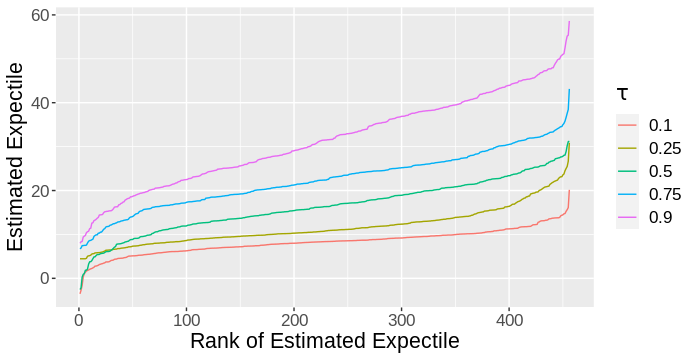}
    \caption{The conditional distribution of smoking quantity for five expectile levels (i.e., 0.1, 0.25, 0.5, 0.75, and 0.9) in gene CHRNB4}
    \label{fig:my_label}
\end{figure}

Table 1, 2, 3 summarize the MSE of ENN with transfer learning and ENN without transfer learning for five different expertiles (i.e., 0.1, 0.25, 0.5, 0.75,
and 0.9). 
Based on the results of  three tables, expectile neural networks with transfer learning outperform expecilt neural networks without transfer learning. To provide a comprehensive view of the conditional distribution of smoking
quantity, we ordered the expectiles estimated based on ENN
from lowest to highest and plotted their values for all
five expectile levels. Fig 3-5 show that the distributions
of estimated expectiles are different across five expectile
levels. When $\tau= 0.5$, ENN models the mean response, in
which the estimated expectiles are closer to its mean. Nonetheless, for high expectile levels (e.g., $\tau$= 0.9),
the estimated expectiles vary among individuals and high-
ranked individuals have much higher expectiles than low-
ranked individuals(e.g., $\tau$= 0.1).

\subsection{Real data application II}
In this real data application, we applied ENN and transfer learning technique to the genetic data from Alzheimer's Disease Neuroimaging Initiative(ADNI). ADNI is a multisite study that aims to improve clinical trials for the prevention and treatment of Alzheimer's disease(AD). APOE allele is the most important genetic risk factor for Alzheimer’s disease\cite{YLL}. We focus our ENN model on APOE gene. After quality control, 165 SNPs remained for the analysis. We only included 677 Caucasian and African America individuals due to the small sample size of other ethnic group. To improve the performance of ENN, we also included 3 covariates: sex(male=1, female=2), age, and education level in the analysis.

Hippocampus is the part of the brain area associated with memories. Alzheimer's 
disease(AD) usually first damages hippocampus, leading to memory loss and 
disorientation. Study showed that hippocampal volume and ratio was reduced by 25\% in 
Alzheimer's disease\cite{AVA}. Hippocampal atrophy and medial temporal lobe cortical 
thickness were associated with the severity of cognitive symptoms\cite{EGJ}. 
Hippocampal atrophy, while not specific for AD, was a fairly sensitive marker of the 
pathologic AD stage and consequent cognitive status\cite{JCR}. Here we used normalized hippocampal volumn as phenotype. The Mini-Mental State 
Examination (MMSE) is a 30-point questionnaire that is used extensively in clinical 
and research settings to measure cognitive impairment. For more information, readers could refer to https://www.ncbi.nlm.nih.gov/projects/gap/cgi-bin/GetPdf.cgi?id=phd001525.1. 80\% of participants score fall into the interval between 27 and 30. Since $\tau$ quantifies location of a distribution, we do not include extreme small ($\tau=0.1$) or larger ($\tau=0.9$) in this analysis. Only three levels of expectile are included in our analysis.

We transfer knowledge learnded from hippocampus-related phenotype to predict score of MMSE. To have stable performance, we also randomly split the dataset 50 times and average the result.

\begin{table}[h]
\caption{The accuracy performance of two models built by ENN with transfer learning(ENN.TF) and ENN in ADNI data} \label{comparsion with two models}
\begin{center}
\begin{tabular}{lllllll}
\hline
       & \multicolumn{2}{c}{\textbf{ENN.TF}}                                       &  & \multicolumn{2}{c}{\textbf{ENN}}    &  \\ \cline{1-3} \cline{5-6}
$\tau$ & \multicolumn{1}{c}{\textbf{Train}} & \multicolumn{1}{c}{\textbf{Test}} &  & \multicolumn{1}{c}{\textbf{Train}} & \multicolumn{1}{c}{\textbf{Test}} &  \\ \hline
0.25   &5.00       &5.17   &     &5.30           &6.78        \\ \hline
0.5    &4.11       &4.31   &     &4.30           &4.82        \\ \hline
0.75   &4.67      &4.88     &    &4.85          &6.86         \\ \hline

\end{tabular}
\end{center}
\end{table}

\begin{figure}[h]
    \centering
    \includegraphics[scale=0.5]{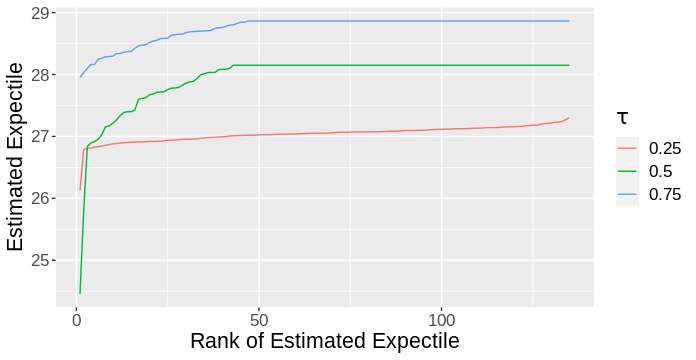}
    \caption{The conditional distribution of smoking quantity for three expectile levels (i.e., 0.25, 0.5 and 0.75) in ADNI}
    \label{fig:my_label}
\end{figure}

From table 4, expectile neural network with transfer learning outperforms expectile regression without transfer learning under different $\tau$. Fig 6 shows that the distributions of estimated vary across three expectile levels. Since the uneven distribution of MMSE score, three sorted expectiles tend to be flat after certain rank. 
\section{Summary and discussion}
From  two real data applications, transfer learning could improve performance of expectile neural networks if it is implemented properly.  In some situations, if the source domain and target domain
are not related to each other or have little relation, brute-force transfer may be unsuccessful. Based on our experience, the outcome of transfer learning relies on what, how and when to transfer. If transfer learning technique is not implemented properly, negative transfer happens where the transfer of knowledge from the source domain to the target domain does not lead to any improvement, but rather causes a drop in the overall performance in the target task.  Negative transfer happens when the source domain/task data
contribute to the reduced performance of learning in the target domain/task. Few research works were proposed on how to overcome negative transfer in the past. Research on how to avoid negative transfer can be further examined in the future.

In genetic data analysis, the sample size for different population is uneven. For example, African American has less samples than Caucasian American.  It is expensive or impossible to collect sufficient training data to train models for
certain rare diseases. It would be more worthwhile if we could reuse the training data when we study certain rare disease for different race. In such cases, Even if the subtle genetic difference among race, transfer learning between source tasks or domains become more desirable and crucial to improve prediction for certain diseases.

  \section*{Acknowledgment}

 The SAGE datasets used for the analyses were obtained from dbGaP at \url{https://www.ncbi.nlm.nih.gov/projects/gap/cgi-bin/study.cgi?study_id=phs000092.v1.p1} through dbGaP accession number phs000092.v1.p1. Alzheimer's Disease Neuroimaging Initiative(ADNI) could be downloaded through \url{http://adni.loni.usc.edu/}.

\ifCLASSOPTIONcaptionsoff
  \newpage
\fi

\end{document}